\documentclass[twocolumn,showpacs,preprintnumbers,amsmath,amssymb]{revtex4}
\usepackage{graphicx}
\usepackage{dcolumn}
\usepackage{bm}

\newcommand{\bef}{\begin{figure}}
\newcommand{\eef}{\end{figure}}

\newcommand{\be}{\begin{equation}}
\newcommand{\ee}{\end{equation}}
\newcommand{\bea}{\begin{eqnarray}}
\newcommand{\eea}{\end{eqnarray}}
\begin{document}

\title{Heavy lepton pair production in nucleus-nucleus collisions at LHC energy - a case study}

\author{Jan-e Alam, Bedangadas Mohanty}
\affiliation{Variable Energy Cyclotron Center, 1/AF, Bidhan Nagar, Kolkata-700064}
\author{Sanjay K Ghosh, Sarbani Majumder, Rajarshi Ray}
\affiliation{Center for Astroparticle Physics \&
Space Science, Block-EN, Sector-V, Salt Lake, Kolkata-700091, INDIA
 \\ \& \\
Department of Physics, Bose Institute, \\
93/1, A. P. C Road, Kolkata - 700009, INDIA}

\date{\today}

\begin{abstract}

We present a study of $\tau^{+}\tau^{-}$ lepton pair production in 
Pb+Pb collisions at $\sqrt{s_{NN}}$ = 5.5 TeV. The larger $\tau^{\pm}$ 
mass ($\sim$ 1.77 GeV) compared to $e^{\pm}$ and $\mu^{\pm}$ leads to 
considerably small hadronic contribution to the $\tau^{+}\tau^{-}$ pair 
invariant mass ($M$)  distribution relative to the production from thermal 
partonic sources.  The quark-anti-quark annihilation processes via 
intermediary virtual photon, Z and Higgs bosons have been considered for the 
production of $ \tau^{+}\tau^{-}$. We observe that the contribution from 
Drell-Yan process dominates over thermal yield  for $\tau^{+}\tau^{-}$ pair 
mass from 4 to 20 GeV  at the LHC energy. We also present the ratio of $\tau$ 
lepton pair yields for nucleus-nucleus collisions relative to yields from p+p 
collisions scaled by number of binary collisions at LHC energies as a function 
$\tau$ pair mass. The ratio is found to be significantly above unity for the 
mass range 4 to 6 GeV. This indicates the possibility of detecting 
$\tau^{+}\tau^{-}$ pair from quark gluon plasma (QGP) in the  mass window 
$4\leq M$(GeV)$\leq 6$. 

\end{abstract}
\pacs{25.75.Ld}
\maketitle

\section{INTRODUCTION}
Dilepton production in high energy heavy-ion collisions have been shown to be an
excellent observable for studying various dynamical aspects of the evolution
of the system formed in heavy-ion collisions~\cite{mclerran,gale,weldon}
(see ~\cite{alam1} for a review). The lepton pair mass distribution
has been used for the  diagnostics of QGP formation as well as 
to study the in-medium properties of low mass vector mesons~\cite{alam2,rapp,BR}. 
The transverse momentum ($p_{\mathrm T}$) distribution of lepton pairs in various $M$
ranges has been used to study the systems radial flow development~\cite{jajati}. 
The HBT interferometry using dilepton pairs has been proposed to provide information on the
time development of collectivity in heavy-ion collisions~\cite{payal}. 

Measurements at SPS (center of mass energy, $\sqrt{s_{NN}}$ = 17.3 GeV)~\cite{ceres,na60} and RHIC 
($\sqrt{s_{NN}}$ = 200 GeV)~\cite{phenix} have provided results for $e^{+}e^{-}$ and $\mu^{+}\mu^{-}$ 
lepton pairs. Hence theoretical calculations have so far also concentrated on 
$e^{+}e^{-}$ and $\mu^{+}\mu^{-}$ dilepton production. With the starting of the 
heavy-ion collision program at Large Hadron Collider (LHC) a value of $\sqrt{s_{\mathrm NN}}=2.76$ TeV
has been achieved for Pb + Pb collisions and 
and it is planned to reach $\sqrt{s_{NN}} = 5.5$ TeV shortly. At the LHC energies we expect 
significant production of the third generation lepton, the $\tau$-leptons. This opens up
the possibility to study $\tau^{+}\tau^{-}$  pair production in addition to 
 $e^{+}e^{-}$ and $\mu^{+}\mu^{-}$ pairs at these energies. The major advantage of 
looking at $\tau^{+}\tau^{-}$ lepton pair arises due to the mass of the  $\tau$ ($\sim$ 1.77 GeV).
The  $\tau$ pair mass distribution would then start beyond the low mass  hadronic
resonances ($\omega$, $\rho$ and $\phi$). That is $\rho$, $\omega$ and $\phi$ can not
''pollute'' (unlike $e^+e^-$ or $\mu^+\mu^-$) the $\tau^+\tau^-$
contributions from QGP in this mass domain, 
making it easier to detect QGP through their measurement. 
The $\tau^+\tau^-$ pair production from the decays of heavy
flavours is expected to be small in the mass region around 4 GeV.
This would in turn mean the remaining contribution
for $\tau$ production are due to thermal partonic sources 
and the Drell Yan (DY). 
Although the yield of $\tau^+\tau^-$
will be lower compared to other lepton pairs, the yield at LHC 
should be significant enough to make their detection possible.

In this paper we present a case study of heavy lepton pair
production for central Pb+Pb collisions at mid-rapidity for $\sqrt{s_{NN}}$ = 5.5 TeV.  
In section II we discuss various processes for $\tau^{+}\tau^{-}$  production.
Section III is dedicated for the space-time 
 description of the evolving thermal medium 
formed in the collision.  Results for thermal as well as DY production are
presented in section IV.  Finally we summarize our results on this case study in 
section V.

\section{SOURCE OF $\tau$ DILEPTON PAIR PRODUCTION}

The main processes for  $\tau^{\pm}$ pair production is by
quark and anti-quark annihilation via intermediary  photon, Z and Higgs bosons. 
The production of $\tau^+\tau^-$ is significant for $M\sim M_Z$ and $M\sim M_H$
for processes mediated by $Z$ and Higgs bosons respectively, for lower $M$ the
contributions from these processes are negligible.
The corresponding Feynman diagrams are shown in Fig.~\ref{fig1}. They all contribute to 
thermal production of $\tau^{\pm}$ pair in quark gluon plasma, in principle as well as
in the DY process. 

The productions for these processes are evaluated from the matrix elements indicated below.
The matrix element for the process $q\bar{q}\rightarrow\tau^+\tau^-$ via $Z$  is given by,
\begin{equation}
 M_Z =\frac{g^2}{4cos^2\theta_w}\frac{1}{(q^2-{m_z}^2)}[\bar{v(p_2)}{\Gamma_q}u(p_1)][\bar{u(k_1)}{\Gamma_\tau}v(k_2)]\\
\label{eqZ}
\end{equation}
with 
\begin{equation*}
 \Gamma_q=\gamma^\mu(c_V^q-c_A^q\gamma_5)\\
\end{equation*}
and
\begin{equation*}
{\Gamma_\tau}=[{\gamma_\mu}-\frac{{q_\mu}{\gamma_\nu}{q^\nu}}{{m_z}^2}][c_V^\tau-c_A^\tau{\gamma_5}]
\end{equation*}
where $\theta$ is the weak mixing angle, $g$ is weak coupling strength, $c_V$'s and $c_A$'s are  
vectors and axial vector couplings~\cite{halzen}. 
The matrix element for the photon mediated process is given by: 
\begin{equation}
M_\gamma=\frac{{e_q}e}{q^2}[\bar{v(p_2)}{\gamma^\mu}{u(p_1)}][\bar{u(k_1)}{\gamma_\mu}{v(k_2)}]\\
\label{eqphoton}
\end{equation}
$e_q$ is the average charge of quarks, e is the electronic charge.
Finally, the matrix element for the Higgs mediated process is:
\begin{equation}
M_H=\frac{{m_q}{m_\tau}}{v^2(s-m_H^2)}[\bar{v(p_2)}u(p_1)][\bar{u(k_1)}v(k_2)]
\label{eqH}
\end{equation}
where ($p_1$, $p_2$) and ($k_1$, $k_2$) are initial state and final state momenta respectively and
$v (\sim 246$ GeV~\cite{halzen}) is the vacuum expectation value of Higgs field.
$m_q$, $m_\tau$, $m_Z$, $m_H$, are the masses of quarks, $\tau$ leptons, $Z$ boson and Higgs respectively.
The total production cross section ($\sigma_q$) 
of $\tau^+\tau^-$ is obtained by taking a coherent sum
of the matrix elements given in Eqs.~\ref{eqZ}, \ref{eqphoton} and
~\ref{eqH} with the following values of various parameters:  
$M_\tau=1.78$ GeV, $m_Z=91$ GeV, $m_H=120$ GeV, $sin\theta_w=0.234$,
$c_A^q=0.5$, $c_V^q=0.19$, $c_A^\tau=-0.5$ and $c_V^\tau=-0.03$.

The production of DY pair from $pp$ collision is obtained by 
folding the partonic cross section ($\sigma_q$) by the parton distribution
functions (PDF) as follows~\cite{ruuskanen}
\bea
\frac{d\sigma^{pp}}{dM^2dy}&=&\frac{1}{N_c}\frac{dx_1}{x_1}\int\frac{dx_2}{x_2}\sum_q\sigma_q
\left[
q(x_1)\bar{q}(x_2)+(1\leftrightarrow 2)\right]\nonumber\\
&&\times\delta(y-\frac{1}{2}ln\frac{x_1}{x_2})
\label{DY}
\eea
where $\sigma_q$ is the partonic cross section,
$q(x_i)$ ($\bar{q}(x_i)$)  is the quark (anti-quark) 
distribution in nucleon, $x_i$ is the Bjorken scaling variable,  $y$ is the
rapidity and $N_c$ is the number of colours.
In the present work CTEQ5M PDF's~\cite{cteq} have been taken 
for the evaluation of the DY contributions. The dilepton yield from 
leading order DY process in 
Pb+Pb collisions is obtained  as follows:
\be
\frac{dN}{dM^2dy}=\frac{N_{\mathrm coll}(b)}{\sigma^{pp}_{\mathrm in}}\times
\frac{d\sigma^{pp}}{dM^2dy}
\label{DYPbPb}
\ee
where $N_{\mathrm coll}(b)$ is the number of binary 
nucleon nucleon collisions
at an impact parameter $b$ calculated using Glauber model~\cite{glauber}
and $\sigma_{\mathrm in}$ is the inelastic
cross section for $pp$ interaction. We have taken $\sigma^{pp}_{\mathrm in}=60$
mb  and $b=3.6$ fm corresponding 
to 0-50\% centrality at $\sqrt{s_{\mathrm NN}}=5.5$ TeV. 
The shadowing of PDF's has been taken from ~\cite{eskola}.

The production of $\tau^+\tau^-$ from a thermally equilibrated 
system of quarks and gluons can be obtained by convolution of the
elementary cross section mentioned above by the thermal
distribution of the quarks participating in the process as follows.
The number of lepton pairs produced per unit volume 
per unit time ($dN/d^4x$) from the annihilation of
thermal quarks is given by:
\begin{equation}
\frac{dN}{d^4x}=\int\frac{d^3p_q}{(2\pi)^3} f_q(p_q)
\frac{d^3p_{\bar{q}}}{(2\pi)^3} f_{\bar{q}}(p_{\bar q})
v_{q\bar{q}}\sigma_q
\label{thermal}
\end{equation}
where $f_q$ ($f_{\bar{q}}$) is the thermal distribution for the 
quark (anti-quark), $v_{q\bar{q}}$ is the relative velocity
between the quark and anti-quark. Making the following change of 
variables one obtains the distribution of $\tau^+\tau^-$ 
in terms of the invariant mass  and momentum ($P$) of the pairs,
\be
\frac{d^3p_q}{E_q}\frac{d^3p_{\bar{q}}}{E_{\bar{q}}} =\pi\frac{d^3P}{E}dM^2
\sqrt{1-\frac{4m_\tau^2}{M^2}}
\label{jacobian}
\ee
where $m_\tau$ is the mass of $\tau$, $E$ is the energy of the pair.
Using Eqs.~\ref{thermal} and \ref{jacobian} and performing the 
space-time integration over the evolution of the thermal system
we get the invariant mass distribution of $\tau^+\tau^-$ at the
mid-rapidity.


\bef
\begin{center}
\includegraphics[scale=0.3]{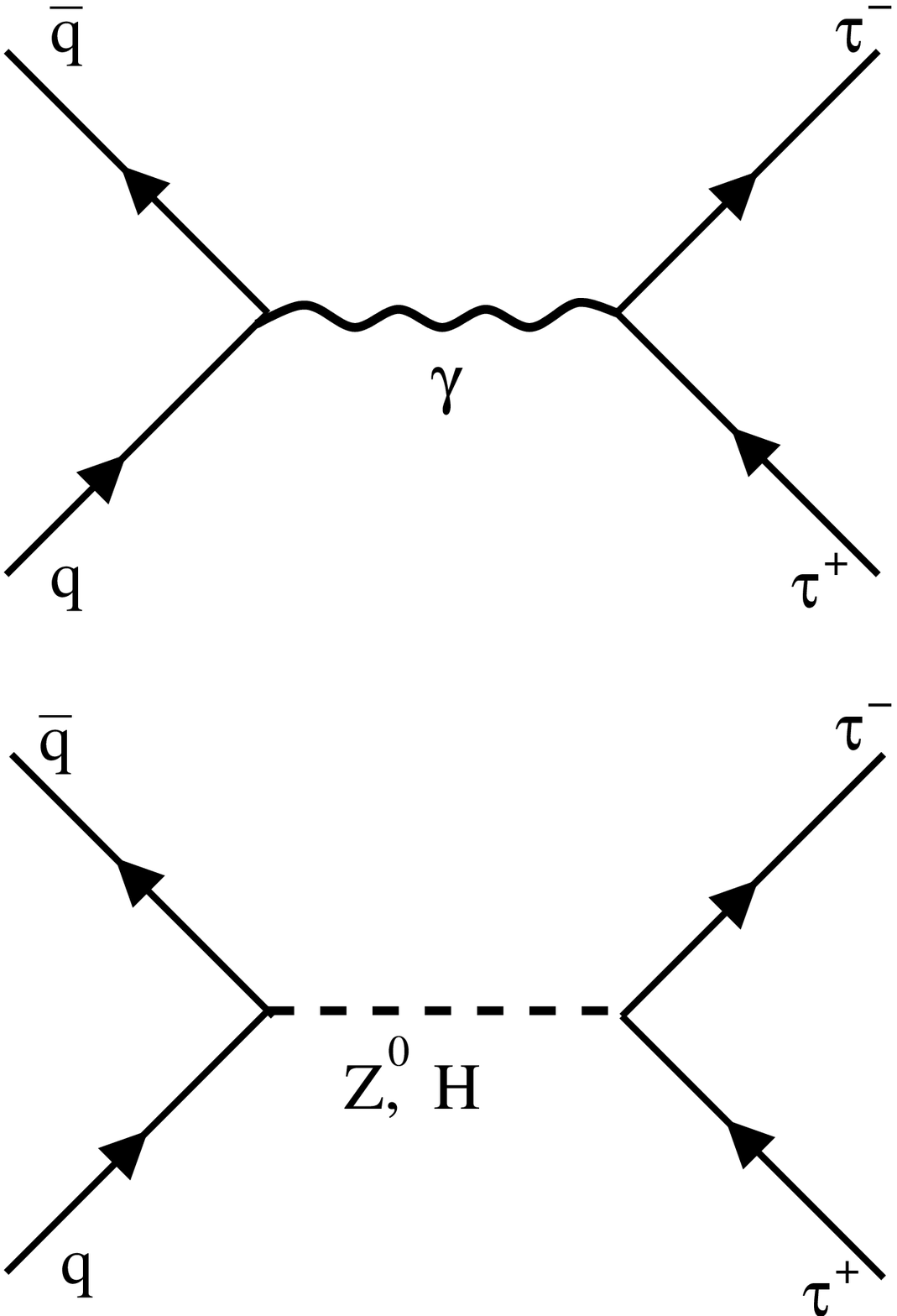}
\caption{Feynman diagrams for heavy dilepton production.}
\label{fig1}
\end{center}
\eef

At multi-TeV energies there could be another contribution to heavy dilepton production
by fusion of two gluons. As discussed in Ref~\cite{willenbrock} the gluon fusion 
process is via a virtual quark loop and an intermediate Z or Higgs boson. This
process was found to be dominant for mass of lepton pair greater than the mass
of W boson. Our results are concentrated in the mass range of 4 to 20 GeV, where
the contribution from such process is negligible.

\section{SPACE-TIME EVOLUTION}

The space time evolution of the system formed
in Pb+Pb collisions at $\sqrt{s_{\mathrm {NN}}}$ = 5.5 TeV 
has been studied by using ideal relativistic hydrodynamics~\cite{ideal} with longitudinal 
boost invariance~\cite{boostinv} and cylindrical symmetry. We assume that the 
system reaches equilibration at a proper time $\tau_{i}$ = 0.08 fm/c after 
the collision. 
The initial temperature, $T_{i}$ is taken to be 700 MeV and is calculated 
assuming the hadronic multiplicity (dN/dy)$\sim 2100$~\cite{armesto}.
We use the equation of state (EoS)  obtained from the lattice QCD
calculations by the MILC collaboration~\cite{milc} for the partonic phase. 
For the hadronic phase EoS all the resonances up to mass 2.5 GeV have been considered~\cite{hadronicph}.
The crossover temperature ($T_{c}$) between hadronic  and partonic phase is taken 
to be 175 MeV~\cite{tc}. 

\section{RESULTS}

\bef
\begin{center}
\includegraphics[scale=0.4]{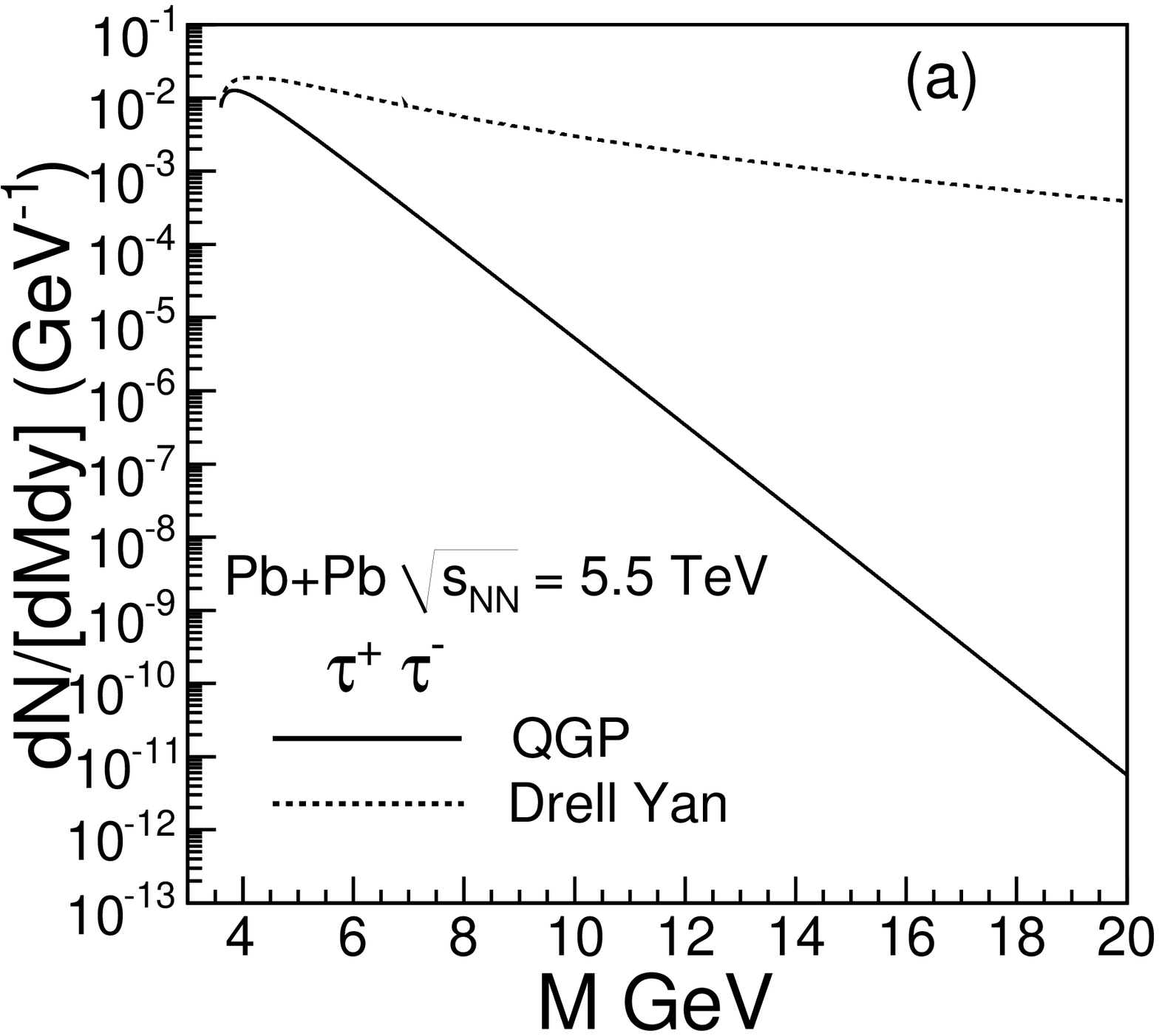}
\includegraphics[scale=0.4]{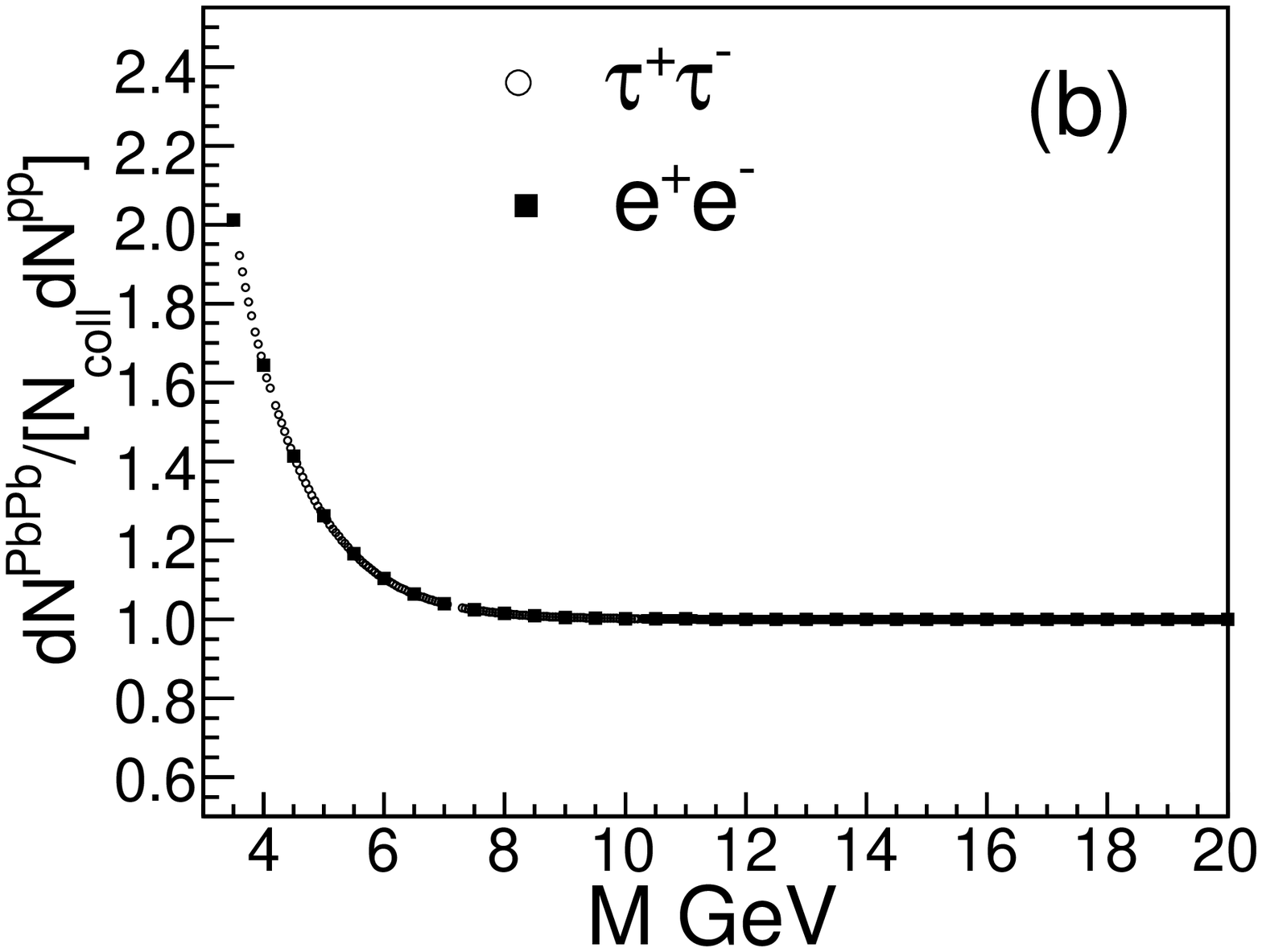}
\caption{(a) $\tau$ lepton pair yields as a function of 
invariant mass of the pair is displayed for  Pb+Pb collisions at $\sqrt{s_{\mathrm {NN}}}$ = 5.5 TeV. 
Solid line indicates the spectra 
from quark gluon plasma and the dashed line stands for  contribution from DY process.
In (b) the ratio $\frac{dN^{PbPb}}{dMdy}$ / $[N_{coll}\frac{dN^{pp}}{dMdy}]$ is shown,
here $\frac{dN^{PbPb}}{dMdy}$ is the sum of the contribution shown in (a), 
$\frac{dN^{pp}}{dMdy}$ is the DY  contribution from $pp$ collision,
and $N_{coll}$ = 1369  for Pb+Pb collisions at $\sqrt{s_{\mathrm {NN}}}$ = 5.5 TeV.
The ratio for electron-positron pair is also displayed (solid square).
}
\label{fig2}
\end{center}
\eef
Figure~\ref{fig2}(a) shows the yield ($\frac{dN}{dMdy}$) for $\tau$ lepton pair
as a function of $\tau^{+}\tau^{-}$ pair invariant mass for Pb+Pb collisions at 
$\sqrt{s_{\mathrm {NN}}}$ = 5.5 TeV.  The contributions from
Drell Yan (DY, dashed line) and thermal partonic medium (QGP, solid line) are shown.
The Drell Yan contribution
is higher than the thermal contribution for all the mass range studied. The difference 
seems to increase with increase in $\tau^{+}\tau^{-}$ pair mass. 

Figure~\ref{fig2}(b) shows the ratio $\frac{dN^{PbPb}}{dMdy}$ / $[N_{coll}\frac{dN^{pp}}{dMdy}]$. 
$\frac{dN^{PbPb}}{dMdy}$ is the sum of the contributions shown in  Figure~\ref{fig2}(a)
from Pb+Pb collisions. The quantity 
$[N_{coll}\frac{dN^{pp}}{dMdy}]$ is the number of binary collisions scaled
contribution from DY process obtained using Eq.~\ref{DYPbPb}. 
This contribution can be estimated from the  measurement 
in p+p collisions at the same energy
($\sqrt{s}$ = 5.5 TeV). We observe that the ratio is above unity for the mass range of 4 to 6 GeV.
Starting with a value $\sim 2$ at mass of 4 GeV it decreases toward unity beyond mass of 6 GeV. 
This indicates that one should be able to extract a clear information of thermal contribution
from partonic source at LHC energies using heavy 
lepton pair measurement within the mass window of 4 to 6 GeV.  It is interesting
to note that the ratio plotted in Fig.~\ref{fig2} for the 
heaviest pairs ($\tau^+\tau^-$) is very similar to that for the 
lightest lepton pairs ($e^+e^-$). 

In this first such case study, we have not discussed the transverse momentum 
distribution of $\tau^{+}\tau^{-}$ pair, these studies are planned to be presented subsequently.
Moreover, we do not discuss the experimental scenario for the 
measurements of  $\tau^+\tau^-$ spectra estimated here 
for  heavy ion collisions at the highest LHC energy. 
For more details we refer
to ~\cite{LHCexpt}.  The $\tau$ particle is the only lepton heavy enough to decay 
into hadrons. $\tau$ leptons are considered to be  a signature in several discovery 
channels related to the Standard Model Higgs boson at low masses,  the MSSM Higgs boson or 
Super-symmetry (SUSY). Hence experimental plans exist at LHC to reconstruct them  
in one-prong (one charged pion) and three-prong (three charged pions) decay topologies.
The lifetime of the $\tau$ lepton ($c\tau$ = 87.11$\mu$m) in principle allows for the reconstruction 
of its decay vertex in the case of three-prong decays. The flight path in the detector increases with 
the Lorentz boost of the $\tau$ lepton, but at the same time the angular separation of the decay 
products decreases. A resulting transverse impact parameter of the $\tau$ decay products can be 
used to distinguish them from objects originating from the production vertex. In fact experiments
at LHC claim the overall efficiency for reconstructing good quality tracks from $\tau$ lepton 
hadronic decays is of the order of  82\%~\cite{LHCexpt}. Further the excellent knowledge of $\tau$ 
decay modes and detection from low energy experiments~\cite{taumeasurement} makes the heavy 
lepton pair production measurements feasible at LHC.

\section{SUMMARY}

We have carried out a first case study of $\tau$ lepton pair production for LHC, 
Pb+Pb collisions at mid-rapidity for $\sqrt{s_{\mathrm {NN}}}$ = 5.5 TeV. The LHC
energy is a factor 27 more compared to RHIC, this should allow for significant
production of $\tau$ leptons for LHC energies. The main sources for  $\tau$ pair 
production  is by quark and anti-quark annihilation 
mediated through photon, Z and Higgs bosons. The contribution from gluon fusion process 
via virtual quark loop and intermediate Z and Higgs boson is negligibly small
in the mass range of our calculations (4 to 20 GeV). 
The Drell Yan contribution is found to be higher than the thermal contribution from partonic
 sources for the entire mass range studied. The non-thermal contributions could be measured 
experimentally through p+p collisions, then the ratio of yields from nucleus-nucleus collisions 
to the yields for the binary collision scaled p+p collisions is found to be above unity for 
the mass range of 4-6 GeV. 
This indicates that the invariant mass window for the observation of thermal 
$\tau$ leptons at LHC energy lies in the domain 4-6 GeV.

\noindent{\bf Acknowledgments}\\
BM is supported by DAE-BRNS project Sanction No.  2010/21/15-BRNS/2026. 
JA is partially supported by DAE-BRNS project Sanction No. 2005/21/5-BRNS/2455.
SM would like to thank CSIR for financial support. We thank Dr Zhangbu Xu
for useful discussions.

\end{document}